\documentclass[12pt]{article}
\usepackage[xdvi]{graphicx}
\usepackage{amssymb}
\newcommand{\bea}{\begin{eqnarray}}
\newcommand{\eea}{\end{eqnarray}}

\makeatletter
\@addtoreset{equation}{section}
\makeatother

\begin{document}
\begin{titlepage}
\begin{flushright}
OU-HET 676/2010
\end{flushright}

\vspace{25ex}

\begin{center}
{\Large\bf
Marginal operators in 
quantum field theory \\
\vspace{1ex}
with extra dimensions
}
\end{center}

\vspace{1ex}

\begin{center}
{\large
Nobuhiro Uekusa
}
\end{center}
\begin{center}
{\it Department of Physics, 
Osaka University \\
Toyonaka, Osaka 560-0043
Japan} \\
\textit{E-mail}: uekusa@het.phys.sci.osaka-u.ac.jp
\end{center}


\vspace{3ex}

\begin{abstract}

The classification of
relevant, marginal and irrelevant operators
is studied in the Randall-Sundrum spacetime.
We find that there exist
marginal and interacting operators
in the Randall-Sundrum spacetime
unlike a higher-dimensional effective theory
near the free-field fixed point.
This gives a direction
to treat 
quantum corrections in the field-theoretical
framework with
extra dimensions by constructing models
out of relevant and marginal operators.

\end{abstract}

\end{titlepage}


\newpage

\section{Introduction}

Physics of extra dimensions 
is an interesting possibility 
of particle physics
beyond the standard model~\cite{Manton:1979kb}-%
\cite{Appelquist:2000nn}.
In the standard model, the framework is based on
a four-dimensional field theory
which is renormalizable
and well-defined including quantum effects.
Once extra dimensions are introduced, 
coupling constants can have negative mass dimensions.
This
gives rise to subsequent divergence
for an infinite number of counterterms. 
At first sight, 
higher-dimensional field theory
seems unsuitable as a framework to include
quantum effects.
Without corrections treated itself, 
it seems mere an effective expression where
all the parameters are
derived from more fundamental theory.

One remarkable feature in the
four-dimensional field theory
is that super-renormal-izable,
renormalizable and non-renormalizable operators
are identified as
relevant, mar-ginal and irrelevant operators.
Renormalizability is not necessarily 
a principle to constrain the theory.
The reason why only 
renormalizable terms are left is because
non-renormalizable operators are negligible at low
energy.
The point to identify the low-energy behavior is 
to treat renormalization group flows by
integrating out the shell of high-momentum degrees 
of freedom.
Indeed, this way 
leads to the appearance of non-renormalizable terms
even if we start with only renormalizable terms.
The classification of relevant,
marginal and irrelevant
operators can be obtained from a distance-rescaling,
whose explicit equations will be given
at the beginning of Section~\ref{sec:scale}.
Although various operators such as
higher-dimension operators inevitably 
arise as quantum effects in theory 
with extra dimensions, 
they may be irrelevant operators that are negligible.
It is straightforward to analyze 
operators for
uncompactified extra dimensions as
a simple extension of this method. 
It is known
that the higher-dimensional effective theory such 
as the higher-dimensional $\phi^4$ theory 
at low energy compared to the cutoff
yields a free-field theory 
unlike the four-dimensional
case with interacting operators.
The idea that relevant and marginal operators 
only have to be in the starting action would
simplify the action. However,
if any higher-dimensional theory is free,
it would lead to a trivial low-energy theory
without other additional interactions
such as brane terms.

It has been suggested that a warped spacetime
tends to differ in running
of couplings from the flat spacetime~\cite{%
Pomarol:2000hp}-%
\cite{Randall:2002qr}.
Because couplings in extra-dimensional theories
have negative mass dimensions, higher-dimension
operators necessarily occur irrespective
of the geometry.
It has been shown that higher-dimension operators generated by loop effects change the values of 
physical quantities~\cite{Uekusa:2009dy}.
Unless higher-dimension operators are taken into
account, the difference between 
the flat and warped spacetimes cannot be concluded.
In addition, the discussion must be
generally made not only for a finite number of
loop calculations.
Following the idea of the four-dimensional 
renormalization group flow, 
all we have to do might be to claim
that the coefficients of higher-dimension operators
are negligible at low energies.
Can we have a nontrivial low-energy theory
with extra dimensions?
It needs to be examined whether
the description of extra dimensions,
interactions and quantum properties
can make sense without requiring an additional
ultraviolet completion.

In this Letter,
we study the issue of the classification of
relevant, marginal and irrelevant operators
in the Randall-Sundrum spacetime.
It is shown that the 
rescaling of distances 
is different between the Randall-Sundrum spacetime
and the flat spacetime.
As a result,
we find marginal operators
quite similar to the four-dimensional case.
These marginal operators are interaction 
operators.
The viewpoint of 
the classified operators gives
a direction to select Lagrangian terms for
model building with extra dimensions.

\section{Scaling in warped space
\label{sec:scale}}

\begin{figure}[htb]
\begin{center}
\includegraphics[width=4cm]{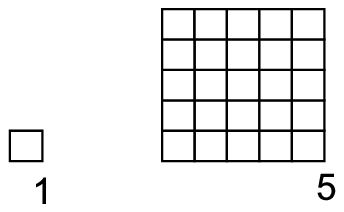} \hspace{10ex}
\includegraphics[width=7cm]{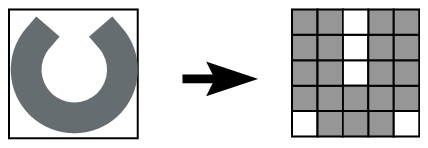}
\caption{The resolution of an 
object in unit square lattice.
For illustration, the figure is drawn 
in two-dimensions.
An intuitive image for the observation
of a picture is depicted.
\label{fig15}}
\end{center}
\end{figure}

\begin{figure}[htb]
\begin{center}
\includegraphics[width=4cm]{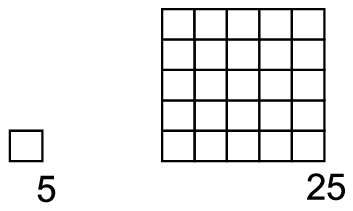}
\hspace{10ex}
\includegraphics[width=7cm]{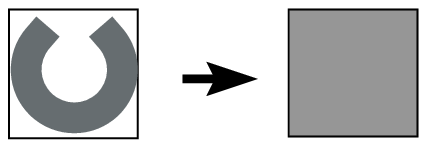}
\caption{The resolution of an 
object for the rescaling $x'=5x$.
\label{fig525}}
\end{center}
\end{figure}

Following Ref.~\cite{PeskinSchroeder},
a convenient way to treat renormalization
group flows is to rescale distances
\bea
   p' = p/b , \qquad x' = x b,
     \label{eqscale}
\eea
where $x$ and $p$ denote 
the four-dimensional coordinates
and the corresponding momenta, respectively
and $b$ is a parameter specifying the scale.
The cutoff $|p|=b \Lambda$ with $b<1$ 
is read in terms of the rescaled momenta as
$|p'|=\Lambda$.
The resolution of
a two-dimensional object of the size $5^2$ by unit square lattice is drawn in Figure~\ref{fig15}.
The object is resolved by $5^2$ squares.
For the rescaling $x'=5x$,
the object is resolved by 1 square
shown in Figure~\ref{fig525}.
The rescaling $x' =5x$ decrease the resolution
by the factor $1/5^2$.
Under the rescaling
(\ref{eqscale}),
the effective action in $\phi^4$ theory,
\bea
   \int d^d x \,
     {\cal L}_{\textrm{\scriptsize eff}}
  =\int d^d x 
   \left[
  {1\over 2}  Z (\partial_\mu \phi)^2
    +{1\over 2} m^2 \phi^2 
     +{1\over 4!}
       \lambda  \phi^4 
          +C (\partial_\mu \phi)^4
            +\ D \phi^6
             \right] ,
\eea
becomes
\bea
     \int d^d  x \,
  {\cal L}_{\textrm{\scriptsize eff}}
  &\!\!\!=\!\!\!&
     \int d^d x' \left[
     {1\over 2} (\partial'_{\mu} \phi')^2
      +{1\over 2} {m'}^2 {\phi'}^2
   + {1\over 4!} \lambda' {\phi'}^4
      + C' (\partial'_{\mu} \phi')^4
     + D' {\phi'}^6  \right] .
      \label{leff2}
\eea
Here $\phi' = \left[ b^{2-d} Z\right]^{1/2}
\phi$,
${m'}^2 =m^2 Z^{-1} b^{-2}$,
$\lambda' = \lambda Z^{-2} b^{d-4}$,
$C' = C Z^{-2} b^d$ and
$D'=D Z^{-3} b^{2d-6}$.
Operators whose coefficients are multiplied by
negative powers of $b$ are relevant operators,
while operators whose coefficients are multiplied by
positive powers of $b$ are irrelevant operators.
For the coefficients of operators
multiplied by $b^0$, the operators are 
marginal operators.
The expressions above have been given
for the dimension of spacetime $d$.
The assumption is
that the extra-dimensional coordinates obey
the same
rescaling as in Eq.~(\ref{eqscale}).
The action integral (\ref{leff2}) shows that
for $d>4$, only the relevant operator is
the mass operator. 
In warped space,
the scaling law is 
different from Eq.~(\ref{eqscale}).
This leads to change of the dependence of operators 
on $b$.

Let us consider the observation of an object
in the Randall-Sundrum spacetime whose
metric is given by \cite{%
Randall:1999ee, 
Randall:1999vf}
\bea
  ds^2 =
    {1\over z^2}
      \left(\eta_{\mu\nu} dx^\mu dx^\nu - {1\over k^2}
      dz^2 \right) ,
      \label{metric}
\eea
where $k$ is the curvature
of the five-dimensional anti-de Sitter space.
Here we examine 
the aspect of the resolution of 
a two-dimensional object
with respect to the extra-dimensional direction.
In the Randall-Sundrum spacetime,
a small $z$ corresponds to a large cutoff.
\begin{figure}[htb]
\begin{center}
\includegraphics[width=6cm]{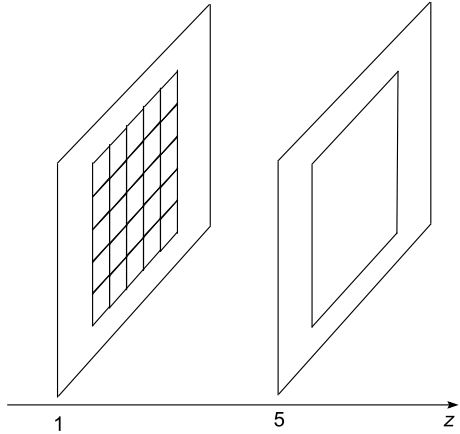}
\caption{The resolution of an object
by a unit square lattice at $z=5$.
\label{fig15z}
}
\end{center}
\end{figure} 
\begin{figure}[htb]
\begin{center}
\includegraphics[width=6cm]{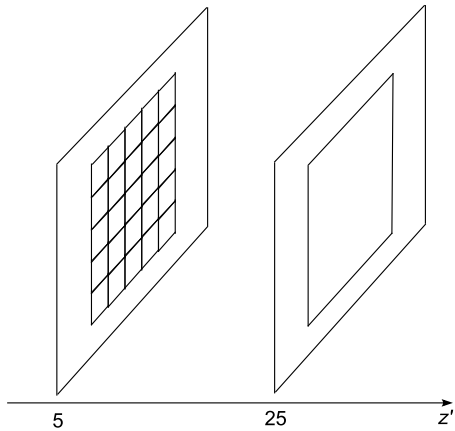}
\caption{The resolution for $z'=5z$.
\label{fig525z}
}
\end{center}
\end{figure}
The resolution of an object 
by 1 square at $z=5$ is
shown in Figure~\ref{fig15z}.
The rescaling $z'=5z$ leads to the change 
shown in Figure~\ref{fig525z}.
For $z'=5$,
the object is resolved by $5^2$ squares.
The rescaling $z'=5z$ increase the resolution
by the factor $5^2$.
This means that the law of the rescaling
is given by
\bea
  p' =p/b ,\qquad x'=bx ,\qquad z'=z/b .
   \label{resrs}
\eea
The behavior of the rescaling (\ref{resrs})
is also seen from that
the metric (\ref{metric}) with $x \to bx$
is identical to the metric (\ref{metric})
with $z\to z/b$.
In the next section,
the rescaling (\ref{resrs}) is
applied to the $\phi^4$ theory.

\section{The operators in $\phi^4$
theory}

We consider
the effective action integral in the $\phi^4$ theory
given by
\bea
  && \int d^4 x dz
     \sqrt{\textrm{det} g_{MN}}
   \left[{1\over 2}Z(\partial_\mu \phi)
     (\partial_\nu \phi) g^{\mu\nu}
    +{1\over 2}
    Z_5 (\partial_z \phi)^2 g^{zz}
    \right.
\nonumber
\\
  && \left.
   +{1\over 2}m^2 \phi^2
   +{1\over 4!} \lambda \phi^4
   + C
   ((\partial_\mu \phi) 
(\partial_\nu \phi) g^{\mu\nu})^2
 +C_5 ((\partial_z \phi)^2 g^{zz})^2
   + D\phi^6 
    \right] ,
\eea
where the capital letters $M,N$ are 
the five-dimensional indices.
Substituting the metric (\ref{metric}) into
the action integral yields
\bea
  &&  \int d^4x {dz\over kz}
  \left[
  {1\over 2 z^2}  Z
  (\partial_\mu \phi)^2
     -{k^2 \over 2 z^2} Z_5 (\partial_z \phi)^2
 \right.
\nonumber
\\
  && \left.
    +{1\over 2 z^4} m^2 \phi^2
    +{1\over 4! z^4} \lambda \phi^4
   +C (\partial_\mu \phi)^4
   +C_5 k^4 (\partial_z \phi)^4
    +{1\over z^4} D \phi^6 \right] .
\eea
Here the abbreviated contraction
stands for
a contraction with $\eta^{\mu\nu}$ such as
$(\partial_{\mu} \phi)^2 
=(\partial_{\mu} \phi)(\partial_{\nu} \phi)
\eta^{\mu\nu}$.
A contraction with $g^{M N}$ is described
in the explicit form with $g^{MN}$ as
$(\partial_{\mu} \phi) 
(\partial_{\nu}\phi) g^{\mu\nu}$.
This rule will be used throughout 
this Letter.
For the rescaling (\ref{resrs}),
the action integral becomes
\bea
   &&  \int d^4x' {dz'\over kz'}
  \left[
  {1\over 2 {z'}^2}  
  (\partial'_\mu \phi')^2
     -{k^2 \over 2 {z'}^2} Z'_5 (\partial'_z \phi')^2
 \right.
\nonumber
\\
  && \left.
    +{1\over 2 {z'}^4} {m'}^2 {\phi'}^2
    +{1\over 4! {z'}^4} \lambda' {\phi'}^4
   +C' (\partial'_\mu \phi')^4
   +C'_5 k^4 (\partial'_z \phi')^4
    +{1\over {z'}^4} D' {\phi'}^6 \right] .
     \label{primephi4}
\eea  
Here quantities with a prime are given by
\bea
  && \phi' = b^{-2} Z^{1/2} \phi ,
   \qquad
   Z'_5 = b^{-4} Z^{-1} Z_5 ,
   \qquad
    {m'}^2 = b^{-4} Z^{-1} m^2 ,
  \qquad
    \lambda'= b^0 Z^{-2} \lambda ,
\nonumber
\\
  &&
     C' = b^{8} Z^{-2} C ,
    \qquad
     C'_5 = b^0 Z^{-2} C_5 ,
    \qquad
      D' = b^4 Z^{-3} D .
\eea
Therefore for Eq.~(\ref{primephi4}),
we find the relevant operators
\bea
    {\phi'}^2  ,\qquad
    (\partial'_z \phi')^2 ,
     \label{phi4re}
\eea
the marginal operators
\bea
    (\partial'_\mu \phi')^2 ,
    \qquad
    {\phi'}^4 ,
    \qquad
    (\partial'_z \phi')^4 ,
     \label{phi4marg}
\eea
and the irrelevant operators
\bea
   (\partial'_\mu \phi')^4 ,
   \qquad
   (\phi')^6  .
     \label{phi4irre}
\eea
The equation~(\ref{phi4marg})
shows that there exist
marginal and interacting operators.
Therefore we can choose the starting action integral
composed of Eqs.~(\ref{phi4re}) and (\ref{phi4marg})
without the irrelevant operators (\ref{phi4irre}).
In general, it is necessary to take
into account more operators.
When the number of constituent
fields is large,
the operators tend to be irrelevant operators.
Since $-g^{zz} \partial_z^2 = k^2z^2 \partial_z^2
=b^0 k^2{z'}^2 {\partial'_z}^2$,
operators with
a large number of $\partial_z$ 
can be marginal operators such as 
the last term in Eq.~(\ref{phi4marg}).
A formally favorable point is 
that the derivative $\partial_z$ corresponds to
picking up the mass. 
Treating these terms
may reduce to analysis for algebraic equations.
The problem of multiplicative $\partial_z$ is found 
also 
for the pure gauge theory in
the next section.

\section{The operators in pure gauge theory}

In this section,
we examine the rescaling behavior of
operators in pure gauge theory 
in the flat spacetime and the Randall-Sundrum
spacetime.

\subsubsection*{Flat spacetime}

In the flat spacetime,
the action integral is given by
\bea
 &&  \int d^d x \,
      \textrm{tr} \left[
        -{1\over 2} F_{\mu\nu} F^{\mu\nu} 
         \right]
\nonumber
\\
  &\!\!\!=\!\!\!&
  \int d^d x \,
   \textrm{tr} \left[
     -{1\over 2}
       (\partial_\mu A_\nu -\partial_\nu A_\mu)^2
   + ig (\partial_\mu A_\nu-\partial_\nu A_\mu) 
     \left[ A^\mu ,A^\nu\right]
     +{1\over 2} g^2 (\left[A_\mu ,A_\nu\right])^2
      \right] ,
\eea
where $F_{\mu\nu} = \partial_\mu A_\nu
-\partial_\nu A_\mu - ig \left[A_\mu ,A_\nu\right]$.
For the rescaling (\ref{eqscale}),
the action integral becomes
\bea
 &&
  \int d^d x \,
   \textrm{tr} \left[
     -{1\over 2}
       (\partial'_\mu A'_\nu -\partial'_\nu A'_\mu)^2
       \right.
\nonumber
\\
  && \left.
   + ig b^{(d-4)/2} (\partial'_\mu A'_\nu
  -\partial'_\nu A'_\mu) 
     \left[ {A'}^\mu ,{A'}^\nu\right]
     +{1\over 2} g^2 
   b^{d-4} (\left[A'_\mu ,A'_\nu\right])^2
      \right] ,
      \label{flatgauge}
\eea
where $A'_\mu =b^{-(d-2)/2} A_\mu$.
From Eq.~(\ref{flatgauge}), it is seen that the
three-point and four-point vertices are 
irrelevant operators for $d>4$.
Thus the bulk action in the flat spacetime
could be taken as a free theory.

\subsubsection*{Randall-Sundrum spacetime}

In the Randall-Sundrum spacetime,
the most simple gauge-field action integral is given by
\bea
  && \int d^4 x dz \,
  \sqrt{\textrm{det}g_{MN}}
   \textrm{tr}\left[
   -{1\over 2}
    F_{MN} F_{N Q} g^{MP} g^{N Q} \right]          
\nonumber
\\
  &\!\!\!=\!\!\!&
    \int d^4 x {dz \over kz} 
   \textrm{tr}
  \left[ -{1\over 2}
  (\partial_\mu A_\nu -\partial_\nu A_\mu)^2
 + ig (\partial_\mu A_\nu -\partial_\nu A_\mu)
  \left[A^\mu , A^\nu\right]
  +{1\over 2} g^2 (\left[A_\mu, A_\nu\right])^2
 \right.
\nonumber
\\
  && \left.
  + k^2 (\partial_\mu A_z -\partial_z A_\mu)^2
  -2i g k^2
  (\partial_\mu A_z -\partial_z A_\mu)
  \left[A^\mu, A_z\right] 
  -g^2 k^2 (\left[A_\mu, A_z \right])^2 
  \right]  .
    \label{rsgauge}
\eea
For the rescaling (\ref{resrs}),
the action integral~(\ref{rsgauge}) becomes
\bea
 &&
    \int d^4 x {dz \over kz} 
   \textrm{tr}
  \left[ -{1\over 2}
  (\partial'_\mu A'_\nu -\partial'_\nu A'_\mu)^2
 + i b^0 g (\partial'_\mu A'_\nu -\partial'_\nu A'_\mu)
  \left[{A'}^\mu , {A'}^\nu\right]
    \right.
\nonumber
\\
 && +{1\over 2} b^0 g^2 (\left[{A'}_\mu, 
   {A'}_\nu\right])^2
  + k^2 (\partial'_\mu {A'}_z - {1\over b^2}
   \partial'_z {A'}_\mu)^2
\nonumber
\\
  && \left.
  -2i b^0 g k^2
  (\partial'_\mu {A'}_z - {1\over b^2}
  \partial'_z {A'}_\mu)
  \left[{A'}^\mu, {A'}_z\right] 
  -b^0 g^2 k^2 (\left[A'_\mu, A'_z \right])^2 
  \right]  .
   \label{rsgaugere}
\eea
where $A'_\mu = A_\mu/b$
and $A'_z = A_z /b$.
From Eq.~(\ref{rsgaugere}), it is found that
the operators with $\partial_z A_\mu$ 
are relevant operators and that
the other terms including the three-point
and four-point vertices are marginal operators.
Particularly the field strength with
four-dimensional indices
obeys the rescaling $F'_{\mu\nu} =F_{\mu\nu}/b^2$.
From the analysis of relevant and
marginal operators up to here,
all the terms in Eq.~(\ref{rsgauge})
need to be included in the starting action integral.
     
To find the rescaling property
of higher-dimension operators,
we write down several explicit examples.
The first example has  the form with
only four-dimensional indices as   
\bea
 &&  \int d^4 x dz \, \sqrt{\textrm{det}
     g_{MN}}
 \textrm{tr}
   \left[ D_\mu F_{\nu \rho} \cdot
       D_{\sigma} F_{\tau \lambda} \cdot
          g^{\mu\nu} g^{\sigma \tau} g^{\rho\lambda}
          \right] 
\nonumber
\\
  &\!\!\!=\!\!\!&
  \int d^4 x {dz \over kz} \,
     z^2
 \textrm{tr}
   \left[ D_\mu F_{\nu \rho} \cdot
       D_{\sigma} F_{\tau \lambda} \cdot
          \eta^{\mu\nu} \eta^{\sigma \tau} 
  \eta^{\rho\lambda}
          \right] ,
          \label{high1}
\eea
where $D_\mu F_{\nu\rho} = \partial_\mu
F_{\nu\rho} -ig \left[A_\mu, F_{\nu\rho}\right]$.
For the rescaling (\ref{resrs}),
Eq.~(\ref{high1})
becomes
\bea
 \int d^4 x' {dz' \over kz'} \,
     {z'}^2 b^4
 \textrm{tr}
   \left[ D'_\mu F'_{\nu \rho} \cdot
       D'_{\sigma} F'_{\tau \lambda} \cdot
          \eta^{\mu\nu} \eta^{\sigma \tau} 
  \eta^{\rho\lambda}
          \right] ,
\eea
where $D'_\mu F'_{\nu\rho} =b^3 D_\mu F_{\nu\rho}$.
This shows that the operator in Eq.~(\ref{high1})
is an irrelevant operator.     
The next example has the form with the
extra-dimensional components,
\bea
 && \int d^4x dz \sqrt{\textrm{det}g_{MN}}
    \textrm{tr}\left[
      D_z F_{z\mu} \cdot
        D_z F_{z\nu} \cdot g^{zz} g^{zz} g^{\mu\nu} 
         \right]
\nonumber
\\
  &\!\!\!=\!\!\!&
  \int d^4 x {dz \over kz} k^4 z^2
   \textrm{tr}
  \left[ (\partial_z F_{z\mu})^2
   -2i g \partial_z F_{z\mu} \cdot
   \left[ A_z , F_z^{\cdot \mu} \right]
  -g^2 (\left[A_z, F_{z\mu}\right])^2 \right] 
  .
\eea
Substituting the rescaling 
(\ref{resrs}) into this equation yields
\bea
   \int d^4 x' {dz'\over kz'}
      k^4 {z'}^2
       \textrm{tr}
     \left[
     (\partial'_z \overline{F}_{z\mu})^2
   -2i g b^2 \partial'_z \overline{F}_{z\mu} \cdot
   \left[ A'_z , \overline{F}_z^{\cdot \mu}\right]
   -g^2 b^4
   (\left[A'_z , \overline{F}_{z\mu}\right])^2
    \right]
   , \label{high2}
\eea
where
$F_{z\mu} = b^2 \overline{F}_{z\mu}$
and $\overline{F}_{z\mu} = b^{-2} \partial'_z
A'_\mu -D'_\mu A'_z$.
In Eq.~(\ref{high2}), it is found that
terms with $\partial'_z$ are relevant or marginal 
operators and that
the other terms without $\partial'_z$
are irrelevant operators.
Even if gauge invariance is required,
a large number of terms with $\partial_z$
are expected as in the $\phi^4$ theory in the 
previous section.
Therefore from the viewpoint of
the classification of operators,
quadratic field strengths (\ref{rsgauge}) 
are necessary for the starting action 
integral in the quantum field theory
and terms with $\partial_z$ must be treated carefully.

\section{The operators in fermionic theory}

In this section, we examine operators for fermions
in the flat spacetime and in the Randall-Sundrum
spacetime. 

\subsubsection*{Flat spacetime}

In the flat spacetime,
the fermionic action integral is written as
\bea
   \int d^d x \, \bar{\Psi}
 i\gamma^\mu (\partial_\mu 
   -i g A_\mu)\Psi .
\eea
For the rescaling (\ref{eqscale}),
this action becomes
\bea
  \int d^d x' \, \bar{\Psi}'
   i\gamma^\mu (\partial'_\mu
   -ig b^{(d-4)/2} A'_\mu ) \Psi' .
\eea 
Here $\Psi' =b^{-(d-1)/2} \Psi$.
For $d>4$, the gauge interaction is an
irrelevant operator.

\subsubsection*{Randall-Sundrum spacetime}

In the Randall-Sundrum spacetime,
the action integral is written as
\bea
   \int d^4 x dz \, \sqrt{\textrm{det}g_{MN}}
  \bar{\Psi} i\Gamma^A e_A^M
    (\partial_M 
    +{1\over 8} \omega_{MBC} [\Gamma^B, \Gamma^C]
       -ig A_M) \Psi ,
\eea
where $e_A^M$ and $\omega_{MBC}$ 
denote the five-dimensional vielbein and 
spin connection, respectively.
This action integral includes the interaction terms
as
\bea
   \int d^4 x {dz\over kz}
   \, {1\over z^3}
    \left[
    \bar{\Psi} i\Gamma^\mu
    (\partial_\mu -ig A_\mu)\Psi
  +g k \bar{\Psi} \Gamma^5 A_z \Psi \right] .
\eea
Substituting the rescaling (\ref{resrs})
into this equation yields
\bea
  \int d^4 x'
    {dz'\over kz'}
    \, {1\over {z'}^3}
    \left[\bar{\Psi}' i\Gamma^\mu
    (\partial'_\mu -i b^0 g A'_\mu)
    \Psi'
    +b^0 g k \bar{\Psi}' \Gamma^5 A'_z \Psi'
   \right] .
\eea
Here $\Psi' =b^{-3}\Psi$.
It is found that
the gauge and effective Yukawa 
interactions of fermions are 
marginal operators.

\section{Conclusion}

We have found relevant and interacting operators
in the Randall-Sundrum spacetime.
In the gauge theory, quadratic terms composed of
the field strength are relevant and marginal operators.
In addition,
it has been shown that the gauge and 
effective Yukawa interactions of fermions
are marginal operators.
This is different from the flat case which has only
irrelevant operators except for the free part.
Hence, interacting theory with extra dimensions
can be treated
in the field-theoretical context.
Our analysis here is general and can be
applied to
various field theories.
We have also found the problem including
a large number of $\partial_z$.
The rescaling behavior of operators
do not seem to constrain the form 
of terms with $\partial_z$.
It may be useful to examine the effect
of multiple $\partial_z$ in explicit models.

As a complementary aspect of the analysis
given here,
the effect of loop corrections of
higher-dimension operators 
with respect to the four-derivative
have been examined in the
flat spacetime.
From the result of a diagram calculation,
the predictability is connected to
a large cutoff compared to the 
compactification scale~\cite{Uekusa:2009dy},
whereas a large cutoff could deteriorate 
the validity of perturbation.
This subtle situation and the free-field behavior
might be a general defect of the 
flat spacetime.

In the Randall-Sundrum spacetime, 
the four-dimensional coordinates are
quite different from the extra-dimensional
coordinate.
We have shown that this changes the quantum aspect
in field theory with extra dimensions. 
To describe particle physics, 
it would be important to formulate
quantum field theory
at the scale much lower than the Planck scale.
In defining the warped space,
the largest cutoff could be   
the intermediate scale
instead of the Planck scale.
In this way, corrections have
an analogous logarithmic
behavior to the four-dimensional 
counterpart~\cite{Uekusa:2010ge}.
It needs to be examined from various viewpoints 
whether a quantum field theory with extra dimensions
can consistently
overcome the problems of the standard model.

\vspace{8ex}

\subsubsection*{Acknowledgments}

This work is supported by Scientific Grants 
from the Ministry of Education
and Science, Grant No.~20244028.







\end{document}